\documentclass[twocolumn,superscriptaddress,prl]{revtex4-1}
\usepackage{mathrsfs,braket}
\usepackage{amssymb, amsbsy, amsmath, latexsym, dsfont, array, layout, graphicx, mathrsfs, color, bm}
\usepackage[normalem]{ulem}

\begin{document}

\title{Bacterial dimensions sensitively regulate surface diffusivity and residence time}

\author{Premkumar Leishangthem}
\affiliation{School of Physics and Optoelectronic Engineering, Hainan University, Haikou 570228, China}
\affiliation{Complex Systems Division, Beijing Computational Science Research Center, Beijing 100193, China}

\author{Xuan Wang}
\affiliation{Department of Physics, Beijing Normal University, Beijing 100875, China}

\author{Junan Chen}
\affiliation{School of Aerospace Engineering, Tsinghua University, Beijing 100084, China}

\author{Shengqi Yang}
\affiliation{School of Aerospace Engineering, Tsinghua University, Beijing 100084, China}

\author{Xinliang Xu}\email{xinliang@hainanu.edu.cn}
\affiliation{School of Physics and Optoelectronic Engineering, Hainan University, Haikou 570228, China}
\affiliation{Complex Systems Division, Beijing Computational Science Research Center, Beijing 100193, China}
\affiliation{Department of Physics, Beijing Normal University, Beijing 100875, China}

\date{\today}

\begin{abstract}
Run-and-tumble is a common but vital strategy that bacteria employ to explore environment suffused with boundaries, as well as to escape from entrapment. In this study we reveal how this strategy and the resulting dynamical behavior can be sensitively regulated by bacterial dimensions. Our results demonstrate that the logarithm of the surface residence time for bacteria with constant tumble bias is linearly related to a dimensionless parameter of bacterial intrinsic size characteristics, where a small variation in bacterial dimensions, which is natural in a suspension, reproduces well the experimentally observed large variation in bacterial residence time. Furthermore, our results predict that the optimal tumble bias corresponding to the maximum surface diffusivity depends strongly on bacterial dimensions, where the same small variation in bacterial dimensions gives rise to a strongly diversified optimal tumble bias and an order of magnitude change in surface diffusivity.
\end{abstract}


\maketitle

The living environment of bacteria is suffused with boundary surfaces \cite{Persat2015}, near which hydrodynamic interactions can trap bacteria in smooth trajectories \cite{Bianchi2017,Cao2022,Leishangthem2024}. Employing a moving strategy that alternates between straight runs and reorientations (tumbles) \cite{Kurzthaler2024}, these microswimmers get to efficiently explore the environment and escape from entrapment \cite{Wadhwa2022, Tian2024}. Tumbling dynamics of bacteria such as {\it Escherichia coli} has been intensively studied \cite{Bianchi2023, Clements2020}, in particularly through two key dynamical properties: the surface diffusivity and the surface residence time because of their direct influence on many important biological processes, e.g., nutrients search \cite{Nasiri2024}, bacterial transport \cite{Wiles2020}, host-tissue invasion \cite{Clements2012}, and biofilm formation \cite{Nadell2016}. An early experiment using digital holographic microscopy suggested that tumbling is significantly suppressed near surface and thus does not provide an effective escaping mechanism \cite{Molaei2014}. In a recent experiment, simultaneous visualization of the body and flagella demonstrates smoking-gun evidence that supports tumbling defined through unbundling of flagella as the dominant escape mechanism \cite{Junot2022}. Interestingly, this experiment observed a large variation in bacterial residence time $\tau$ spanning over two orders of magnitude, where the distribution of $\ln\tau$ fits well to a Gaussian. In the meantime, while bacterial diffusion along the boundary surface depends strongly on the frequency of tumbling events, recent experimental studies showed that the tumble bias (the ratio of time during tumbles to the total time) for wild-type {\it E. coli} is at the optimal value that the corresponding bacterial surface diffusivity is at maximum \cite{Ipina2019, JunhuaYuan2024}. However, the fundamental mechanism underlying these observations is still unclear, as existing models \cite{Drescher2011, Junot2022, Ipina2019, JunhuaYuan2024} remain phenomenological using experimental observables as parameters (e.g., bacterial velocity $u$ and trajectory radius of curvature $R_\mathrm{c}$). Here we provide a quantitative study of the hydrodynamics involved in both the run and the tumble behavioural states, through numerical simulations based on a simplified model of {\it E. coli} \cite{Zhang2021, Leishangthem2024}. Our results demonstrate that the phenomenological parameters used in previous models depends strongly on bacterial dimensions, where the latter being the fundamental variables that enable efficient regulation of bacterial surface residence and exploration behavior over a broad range. Specifically, we demonstrate that strong diversity in both bacterial residence time and surface diffusivity (and the corresponding optimal tumble bias) can be induced by small variations of bacterial dimensions that are natural in any suspensions, a discovery with both ecological \cite{Keegstra2022} and engineering \cite{Liu2016, Ketzetzi2020} implications.

\begin{figure}[b]
\includegraphics[width=0.45\textwidth]{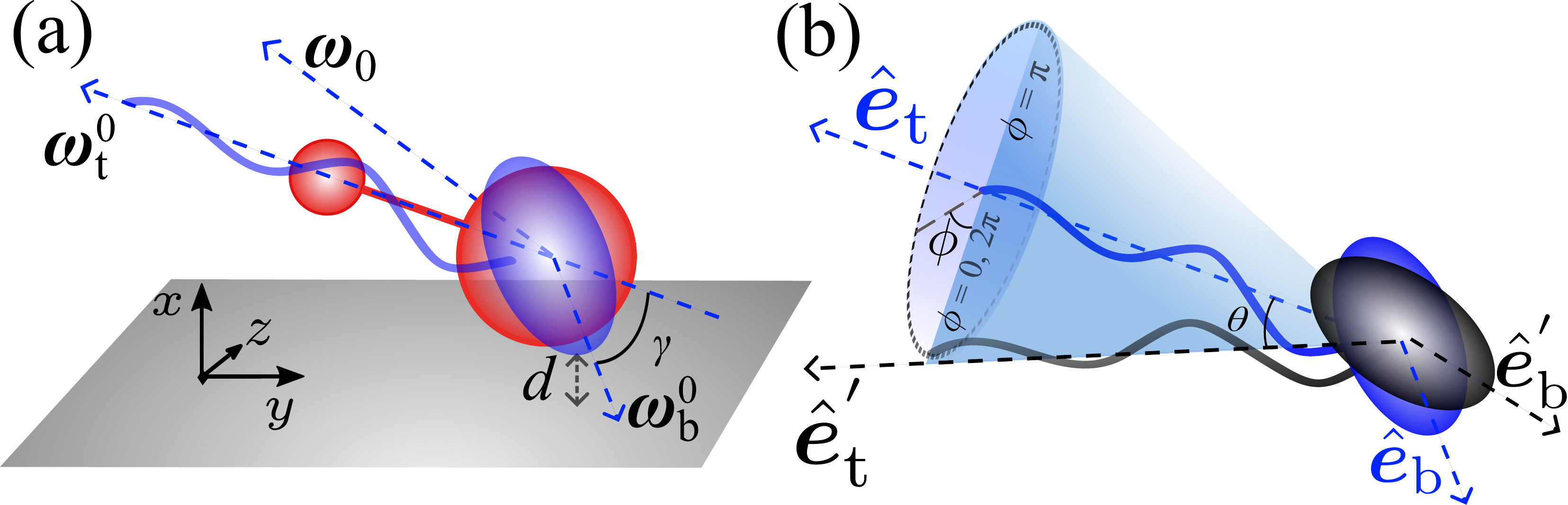}
\caption{(a) An {\it E. coli} (blue) is modeled as two spheres connected by a rod (red). (b) Tumbling in our model is treated as a reorientation from the blue bacterium to the gray bacterium, characterized by $\theta$ and $\phi$.}\label{Fig1}
\end{figure}

\begin{figure*}[ht]
\includegraphics[width=0.85\textwidth]{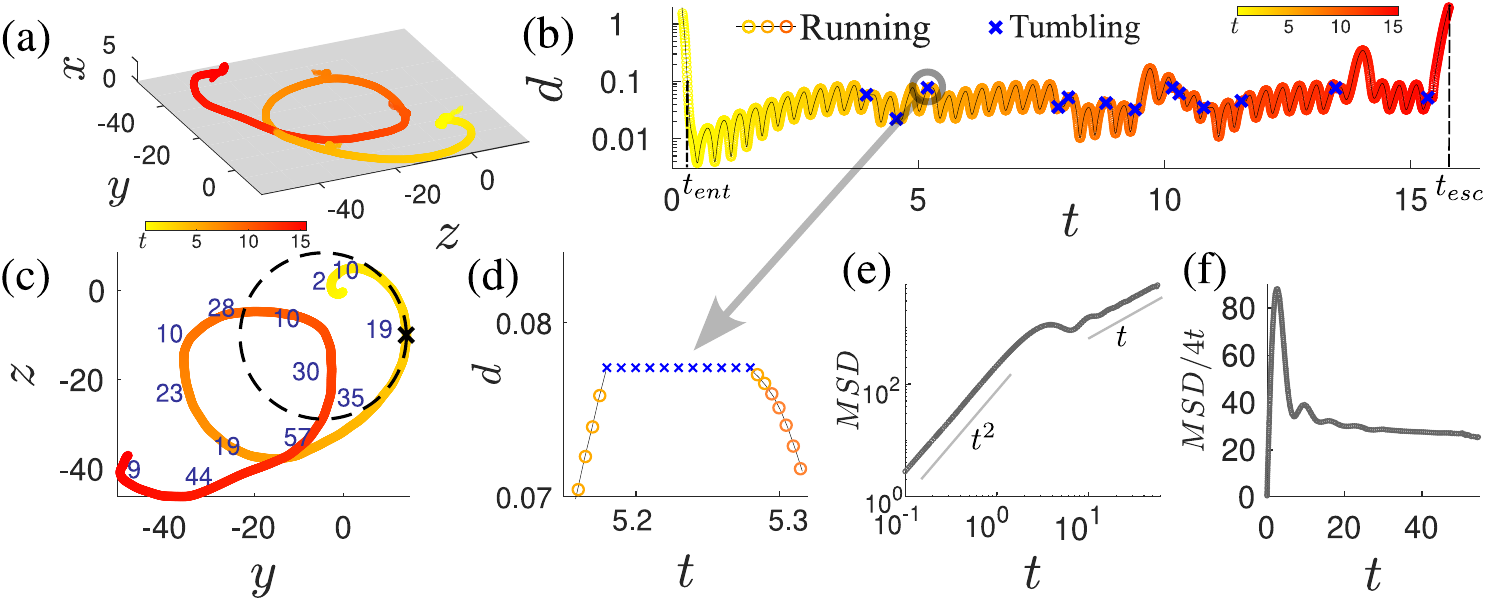}
\caption{Bacterial trajectory in 3D (a), and the corresponding dynamics in $x$ direction perpendicular to the wall (b) and in the 2D $yz$ plane parallel to the wall (c), color coded by time $t$. Tumbling events are marked in (b) as blue crosses, during which surface distance $d$ between body-bead and the wall remains a constant (d). Using the trajectory in the 2D $yz$ plane, mean square displacement $MSD$ (e) and $MSD/4t$ (f) are obtained as functions of time $t$.}\label{Fig2}
\end{figure*}

\textit{Bacteria model and simulation methods--} We study the dynamics of one {\it E. coli} bacterium, which swims in water above an infinitely large plane with no-slip boundary at $x=0$ (Fig.\,\ref{Fig1}a). The Reynolds number of the system is low ($10^{-5}$) so that bacterial flows are typically studied by the linear Stokes equation. To evaluate the hydrodynamic interactions in this regime, bacterial body and flagellar bundle can be simplified as two spherical beads: a body-bead with radius $R_\mathrm{b}$ and a tail-bead with radius $R_\mathrm{t}$, separated in a center-to-center distance of $l$ (Fig.\,\ref{Fig1}a) \cite{Zhang2021}. For simplicity, we assume that the bacterium has only two behavioural states: swimming and tumbling, and the transition between the two states are governed by rates $k_{st}$ and $k_{ts}$ where tumble bias is defined as $TB\equiv k_{st}/(k_{st}+k_{ts})$. For wild type {\it E. coli} bacteria, we have $k_{st}\approx 1 \mathrm{s}^{-1}$ and $k_{ts}\approx 10 \mathrm{s}^{-1}$ \cite{HBergBook}, corresponding to an average running time of $1 s$ between two tumbling events and a mean tumbling time of $0.1 \mathrm{s}$.

When the bacterium is in swimming state, we simulate the dynamics with a time resolution $\Delta t = 10^{-2} s$ where the system is in the overdamped limit described by
\begin{equation}\label{Langevin}
\bm{\mathrm{\xi}}\cdot(\bm{\mathrm{U}}-k_\mathrm{B}T\mathbf{\nabla}\cdot\bm{\mathrm{\xi}}^{-1})=\bm{\mathrm{F}}^\mathrm{P}+\bm{\mathrm{F}}^\mathrm{B}.
\end{equation}
Here the resistance tensor $\bm{\mathrm{\xi}}$ for any given configuration is fully determined from hydrodynamics, $k_\mathrm{B}T\mathbf{\nabla}\cdot\bm{\mathrm{\xi}}^{-1}$ is the thermodynamic effect that becomes significant near the boundary \cite{Leishangthem2024}, $\bm{\mathrm{U}}\equiv\left(\bm{u}_\mathrm{b}, \bm{\omega}_\mathrm{b}, \bm{u}_\mathrm{t}, \bm{\omega}_\mathrm{t}\right)^T$ is the translational/rotational velocity vector characterizing motions of the body-bead and tail-bead, respectively; $\bm{\mathrm{F}}^\mathrm{P}\equiv\left(\bm{F}_\mathrm{b}, \bm{L}_\mathrm{b}, \bm{F}_\mathrm{t}, \bm{L}_\mathrm{t}\right)^T$ represents the nonhydrodynamic forces, and $\bm{\mathrm{F}}^\mathrm{B}$ represents stochastic forces characterized by $\langle\bm{\mathrm{F}}^\mathrm{B}(\Delta t)\rangle=0$ and $\langle\bm{\mathrm{F}}^\mathrm{B}(\Delta t)\bm{\mathrm{F}}^\mathrm{B}(\Delta t)\rangle=2k_\mathrm{B}T\bm{\mathrm{\xi}}\Delta t$. The system dynamics can be resolved with the application of two widely used conditions. One is the free-swimming condition, in the form of $\bm{F}_\mathrm{t}=-\bm{F}_\mathrm{b}\equiv-\bm{F}_\mathrm{eff}$, $\bm{L}_\mathrm{b}=-\bm{L}_\mathrm{t}-(\bm{r}_\mathrm{t}-\bm{r}_\mathrm{b} )\times\bm{F}_\mathrm{t}\equiv-\bm{L}_\mathrm{eff}+\bm{F}_\mathrm{eff}\times(\bm{r}_\mathrm{b}-\bm{r}_\mathrm{t})$. The other condition concerns the connection between the body and the flagellar bundle, where self-spin of bacterial body $\bm{\omega}_b^0$ along its major axis $\hat{\bm{e}}_\mathrm{b}$ and self-spin of the flagellar bundle $\bm{\omega}_t^0$ along its longitudinal axis $\hat{\bm{e}}_\mathrm{t}$, driven by the rotary motor, are generally not collinear. The nonzero angle $\gamma$ formed by the two vectors leads to a center of mass rotation $\bm{\omega}_0$ for free-swimming bacteria and therefore, bacterial wobbling (Fig.\,\ref{Fig1}a). In our model we simplified the connection as a universal joint, which requires $\bm{\omega}_\mathrm{b}-\bm{\omega}_\mathrm{b}^0=\bm{\omega}_\mathrm{t}-\bm{\omega}_\mathrm{t}^0\equiv\bm{\omega}_0$ and $\bm{u}_\mathrm{t}=\bm{u}_\mathrm{b}+\bm{\omega}_0\times(\bm{r}_\mathrm{t}-\bm{r}_\mathrm{b})$, where self-spinning rotations $\bm{\omega}_\mathrm{b}^0$ and $\bm{\omega}_\mathrm{t}^0$ satisfy $R_\mathrm{b}^3|\bm{\omega}_\mathrm{b}^0|\cos{\gamma}=R_\mathrm{t}^3|\bm{\omega}_\mathrm{t}^0|$ \cite{Kamdar2022} with $|\bm{\omega}_\mathrm{t}^0|=\mathrm{100 Hz}$ fixed to match the experiments \cite{Darnton2007}.

The tumbling event of an {\it E. coli} bacterium is widely believed to be triggered by the reverse of the rotational direction of one or more filaments leading to the separation of them from the flagellar bundle, and ended with the reassembly of the bundle as all filaments resume the original rotational direction \cite{HBergBook}. The most significant change during the process, as observed in experiments \cite{Turner2000}, is a reorientation of the bacterium about the center of bacterial body characterized by a polar angle $\theta$ and an azimuthal angle $\phi$ (Fig.\,\ref{Fig1}b). Previous experiments \cite{Turner2000,Taute2015} show that in bulk suspension, $\theta$ has a mean of about 1 $\mathrm{rad}$ and $\phi$ is random between 0 and $2\pi$. For bacteria near boundaries, data about $\theta$ and $\phi$ are scarce. Here in our model we assume that $\phi$ is still random between 0 and $2\pi$, while $\theta$ is affected by the nearby boundary that it is inversely proportional to the rotational friction.

\textit{Numerical simulation with fixed tumble bias--} For a typical {\it E. coli} in water at room temperature, we estimate that $k_\mathrm{B}T \approx 4\times 10^{-21} \mathrm{N \cdot m}$, $\mu\approx 10^{-3} \mathrm{N \cdot s/m^2}$, $R_\mathrm{b} \approx 1 \mathrm{\mu m}$, $l \approx 5 \mathrm{\mu m}$, $R_\mathrm{t} \approx 0.4 \mathrm{\mu m}$, and $\gamma = 30^\circ$ \cite{Darnton2007,Dvoriashyna2021,Kamdar2022}. Using $1 \mathrm{\mu m}$ and $1 \mathrm{s}$ as the unit of length and time respectively and setting $\mu=1$ for the unit of force, such a bacterium is characterized by $\{l=5,R_\mathrm{b}=1,R_\mathrm{t}=0.4,k_\mathrm{B}T=4\}$, for which we simulate 500 independent trajectories with the same fixed initial condition swimming towards the plane. For each trajectory, the bacterium can later randomly switch from swimming state to tumbling state with a rate of $k_{st}=1\mathrm{s}^{-1}$, and then randomly switch back with a rate of $k_{ts}=10\mathrm{s}^{-1}$. During the tumbling state, the bacterium simply reorients without any translation, following our description in previous paragraph. In swimming state, the evolution of the system is simulated using Eq.\,\ref{Langevin}, where the key is getting $\bm{\mathrm{\xi}}$ for each configuration by a two-step procedure following the Stokesian dynamics simulation \cite{Brady1988,Swan2007}. Specifically, we first obtain the far-field mobility tensor through $\bm{M}_{\infty}=\bm{M}_{0}+\bm{\hat{M}}$. Here $\bm{M}_{0}$ is the analytic far-field hydrodynamic interaction without plane \cite{Jeffrey1984}, and the plane contribution $\bm{\hat{M}}$ is analytically available through the method of images \cite{Swan2007}. In the second step, $\bm{\xi}$ is obtained through $\bm{\xi}=\bm{M}_{\infty}^{-1}+\bm{\xi}_\mathrm{b}$, with $\bm{\xi}_\mathrm{b}$ the analytic lubrication between body-sphere and the plane \cite{Bossis1991, Goldman1967}.

In Fig.\,\ref{Fig2}a we show a typical bacterial trajectory, where we highlight the self-spinning by drawing an ellipsoid and a helix in place of the body sphere and tail sphere used in our actual simulation. In Fig.\,\ref{Fig2}b and Fig.\,\ref{Fig2}c we show the corresponding dynamics in $x$ direction perpendicular to the wall and in the $yz$ plane parallel to the wall, respectively, where tumbling events are highlighted in Fig.\,\ref{Fig2}b and Fig.\,\ref{Fig2}d as blue crosses. From the temporal evolution in $x$ (Fig.\,\ref{Fig2}b) two important points in time can be defined: $t=t_{ent}$ the starting time of bacterial entrapment defined as the earliest time when surface distance $d$ between body bead and the plane (Fig.\,\ref{Fig1}a) reaches our set threshold $d=0.1$; and $t=t_{esc}$ the time the bacterium escapes from entrapment characterized by a surface distance $d=2$. The surface residence time can then be obtained as $\tau\equiv t_{esc}-t_{ent}$. From Fig.\,\ref{Fig2}c that shows the bacterial dynamics in the $yz$ plane, we can obtain the local radius of curvature $R_\mathrm{c}$ of the trajectory, mean square displacement ($MSD$) along the surface (Fig.\,\ref{Fig2}e) and the associated diffusion coefficient $D\equiv \lim_{t\to\infty} MSD/4t$ (Fig.\,\ref{Fig2}f). By averaging results from the 500 statistically independent simulation runs, we get $\ln \tau = 2.6$, $R_\mathrm{c} = 20$, and $D = 30$ in agreement with experimentally obtained $\ln \tau=2.39$ \cite{Junot2022}, $10<R_\mathrm{c}<50$ \cite{Baillou2023}, and $10<D<100$ \cite{JunhuaYuan2024}, where the units for time and length are $1 \mathrm{s}$ and $1 \mathrm{\mu m}$, respectively.

\begin{figure}[t]
\includegraphics[width=0.45\textwidth]{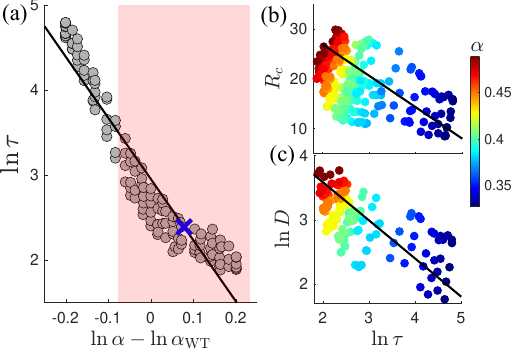}
\caption{Dynamics for bacteria with fixed tumble bias. (a) $\ln \tau$ as a function of $\ln \alpha - \ln \alpha_\mathrm{WT}$ where $\alpha\equiv R_\mathrm{t}/R_\mathrm{b}$, obtained from simulation (gray circles). The black line is a linear fit. The normal distributed $\ln \tau$ observed in experiment \cite{Junot2022} corresponds to $\ln \alpha - \ln \alpha_\mathrm{WT}\approx 0.08 \pm 0.15$ (blue cross for the mean, and red shaded area covers one standard deviation). Scatter plots of $\{R_\mathrm{c}, \ln \tau \}$ (b) and $\{\ln D, \ln \tau \}$ (c) are obtained from simulation (symbols) where the color for each symbol denotes the corresponding $\alpha$. Black lines in (b) and (c) are both linear fit of the data.}\label{Fig3}
\end{figure}

A variety of bacteria characterized by $\{l,R_\mathrm{b},R_\mathrm{t}\}$ are then simulated in a similar fashion, with temperature fixed at $300\mathrm{K}$ and $\gamma$ fixed at $30^\circ$. Here for each of the three parameters we limit the variations to at most $10\%$ from the corresponding value of a typical {\it E. coli}. Our results show surface residence times ranging from $5\mathrm{s}$ to $200\mathrm{s}$, which strongly depends on one dimensionless parameter of bacterial intrinsic size characteristics $\alpha\equiv R_\mathrm{t}/R_\mathrm{b}$, with $\alpha_\mathrm{WT}=0.4$ for a typical bacterium. Specifically, our data fits well linearly in a form of $\ln \tau = -18\alpha + 10$ or equivalently $\ln \tau = -7.2(\ln \alpha - \ln \alpha_\mathrm{WT}) + 3.0$ (solid line in Fig.\,\ref{Fig3}a), where we use $\ln \alpha - \ln \alpha_\mathrm{WT} \approx \alpha/\alpha_\mathrm{WT} - 1$ since variation in $\alpha$ is small. Phenomenologically, the orientational dynamics of a bacterium near a surface is described by a Langevin equation $\dot{\theta}=-dU/d\theta+\sqrt{2D_r}\zeta(t)$ \cite{Drescher2011}, where $U(\theta)=\theta^2/2\kappa$ is an effective potential with time scale $\kappa$ characterizing hydrodynamic realignment, $D_r$ is the effective rotational diffusion coefficient which is dominated by tumbling here, and $\zeta(t)$ is a white noise with unit standard deviation. This is a classical Kramer problem about a noise-induced escape from a potential, with residence time $\ln\tau \propto \theta_\mathrm{esc}^2/(2\kappa D_r)$, where $\theta_\mathrm{esc}$, the minimum reorientation required for escape, can be treated as a constant \cite{Leishangthem2024}. The two phenomenological parameters used, $\kappa$ and $D_r$, depends strongly on bacterial dimensions in the form of $\kappa \propto R_\mathrm{b}\ln(d/R_\mathrm{b})/R_\mathrm{t}$ and $D_r \propto (\ln(d/R_\mathrm{b}))^{-1}$ \cite{Leishangthem2024}, leading to $\ln \tau \propto R_\mathrm{t}/R_\mathrm{b}\equiv \alpha$.

According to this linear relation, the experimentally observed large variation in $\tau$, i.e. the Gaussian distributed $\ln \tau$ with a mean of $2.39$ and a standard deviation of $1.12$ \cite{Junot2022}, corresponds to a Gaussian distributed $\alpha$ with $\alpha/\alpha_\mathrm{WT} - 1 = 0.08 \pm 0.15$ (blue cross and red shaded area in Fig.\,\ref{Fig3}a). In other words, the most typical $\alpha$ estimated from the residence time measurement is only $8\%$ different from the $\alpha_\mathrm{WT}=0.4$ extracted from other independent measurements \cite{Kamdar2022, Darnton2007, Dvoriashyna2021}, and the standard deviation in $\alpha$ estimated from the residence time measurement is $15\%$ of the mean, which is a reasonable presentation of the natural variation of $\alpha$ in an $\it E. coli$ suspension. Furthermore, when plotting dynamical quantities readily accessible in experiments, our results predict strong anticorrelations between $\ln \tau$ and $R_\mathrm{c}$ that fits to $R_\mathrm{c} = -6 \ln \tau + 40$ (Fig.\,\ref{Fig3}b), and between $\ln \tau$ and $\ln D$ that fits to $\ln D = -0.5 \ln \tau + 5$ (Fig.\,\ref{Fig3}c), where the units for time and length are $1\mathrm{s}$ and $1\mathrm{\mu m}$, respectively.

In addition, our simulation reproduces two experimental observations that demonstrates tumbling as the dominate escape mechanism \cite{Junot2022}: simulation runs with tumbling show that the time between escape and the last tumbling event before escape $t_{last}$ is much smaller than the average time between tumbling events $k_{st}^{-1}$ (SI, Sec.\,A); and simulation runs with tumbling state turned off show significantly longer residence times (SI, Sec.\,B).

\begin{figure}[t]
\includegraphics[width=0.45\textwidth]{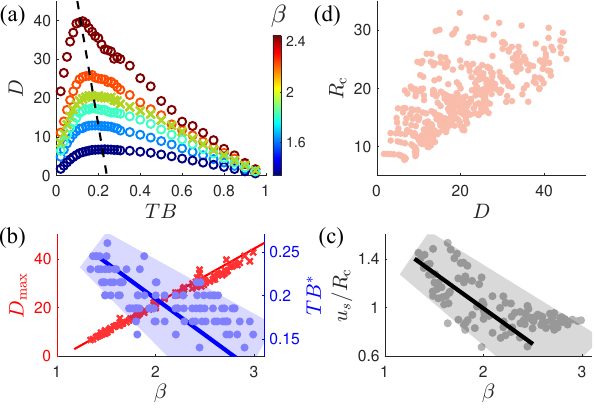}
\caption{Optimal transport along surface. (a) Simulation data for surface diffusivity $D$ as a function of tumble bias $TB$, for different bacteria, color coded by the corresponding $\beta\equiv R_\mathrm{t}l/R^2_\mathrm{b}$. The black dashed line is a guide that connects the optimal transport location $\{TB^*, D_\mathrm{max}\}$ of each curve. (b) $TB^*$ (blue) and $D_\mathrm{max}$ (red) from simulations (symbols) as functions of $\beta$. The red line is a linear fit, while the blue line and blue shaded area are obtained through phenomenological model with $D_s=5 \mathrm{s}^{-1}$ and $\Omega=22 \mathrm{s}^{-1}$. (c) $u_s/R_\mathrm{c}$ from simulation data (gray circles). The black line is a linear fit and the gray shaded area covers the broad distribution. (d) Scatter plot of $\{R_\mathrm{c}, D\}$ from simulation demonstrates a positive correlation.}\label{Fig4}
\end{figure}

\textit{Optimal transport along surfaces --} In this section we study bacterial translation along the surface at different tumble bias $TB$. According to a recent model generalized from experimental data \cite{JunhuaYuan2024}, the transition rates $k_{st}$ and $k_{ts}$ are simple functions of $TB$ and the so-called switching frequency $\Omega$ in the form of $k_{st}\equiv \Omega TB/2$ and $k_{ts}\equiv \Omega (1-TB)/2$. Here we first use $k_{st}^\mathrm{WT}=1\mathrm{s}^{-1}$ and $k_{ts}^\mathrm{WT}=10\mathrm{s}^{-1}$ for the wild type to get $\Omega = 2(k_{st}^\mathrm{WT}+k_{ts}^\mathrm{WT}) = 22\mathrm{s}^{-1}$. Then we treat $\Omega$ as a constant and study bacterial dynamics at different $TB$. Using the typical bacterium with $\{l=5,R_\mathrm{b}=1,R_\mathrm{t}=0.4\}$ as an example, at a given $TB$ we can obtain $k_{st}$ and $k_{ts}$, with which we simulate system dynamics for 500 independent runs and obtain surface translational diffusivity $D\equiv \lim_{t\to\infty} MSD/4t$ as the average. The results of $D$ at different $TB$ are illustrated as crosses in Fig.\,\ref{Fig4}a, which clearly demonstrate the existence of an optimal tumble bias $TB^*=0.18$ where surface diffusivity is at maximum $D_\mathrm{max}$, in qualitative agreement with previous experimental results of $TB^*\approx 0.2$ \cite{Ipina2019, JunhuaYuan2024}. The existence of $TB^*$ arises since $D$ is small at both of the following two limiting cases: at $TB=0$ the deviation from otherwise stable circular trajectory can only arise from thermal noise which is much weaker than tumbling; and at $TB=1$ the bacterium barely moves along the surface as it keeps reorienting without swimming. This existence of $TB^*$ is further validated by simulations of all other bacteria in the same set described in previous section with different choices of $\{l,R_\mathrm{b},R_\mathrm{t}\}$, where some examples are illustrated in Fig.\,\ref{Fig4}a. Extracting $\{TB^*, D_\mathrm{max}\}$ from each curve, our results for all bacteria studied show that both $TB^*$ and $D_\mathrm{max}$ depend strongly on dimensionless parameter of bacterial intrinsic sizes $\beta\equiv R_\mathrm{t}l/R_\mathrm{b}^2$ ($\beta_\mathrm{WT}=2$ for a typical bacterium), as illustrated in Fig.\,\ref{Fig4}b. These results demonstrate a well-defined linear relation between $D_\mathrm{max}$ and $\beta$ (red crosses), where the same small variation in bacterial dimensions leads to a 10-fold change in $D_\mathrm{max}$. In the meantime, $TB^*$ (blue circles) shows a broadly distributed anticorrelation with $\beta$.

We believe this broad anticorrelation arises from a broad distribution of $u_s/R_\mathrm{c}$ in simulation as $\beta$ changes, as illustrated in Fig.\,\ref{Fig4}c where the black line shows a linear fit. Existing models simplified $D$ as a function of a few phenomenological parameters including $TB$, $\Omega$, $u_s/R_\mathrm{c}$, $D_s$, etc (Eq. 7 in \cite{Ipina2019} and Eq. 9 in \cite{JunhuaYuan2024}), where $u_s$ and $D_s$ are velocity and angular diffusivity in the $yz$ plane parallel to the surface, respectively. Through $\partial D/\partial TB=0$, $TB^*$ can be obtained, which reduces to $A_4(TB^*)^4+A_3(TB^*)^3+A_2(TB^*)^2+A_1TB^*+A_0=0$. Here all coefficients $A_i$ are functions of $D_s$, $u_s/R_\mathrm{c}$, and $\Omega$: $A_0=(D_s+\Omega/2)^3 u^2_s/R^2_\mathrm{c}$, $A_1=-2(D_s+\Omega/2)^2(D_s+\Omega)u^2_s/R^2_\mathrm{c}$, $A_2=-\Omega(D_s+\Omega/2)(-5D_su^2_s/R^2_\mathrm{c}+D^2_s\Omega/2-3\Omega u^2_s/R^2_\mathrm{c})/2$, $A_3=-\Omega^2 (D_s+\Omega/2) u^2_s/R^2_\mathrm{c}$, and $A_4=(D^2_s+u^2_s/R^2_\mathrm{c})\Omega^3/8$. With $D_s = 5$ and $\Omega=22$ fixed, $TB^*$ numerically obtained from this phenomenological equation is a monotonous function of $u_s/R_\mathrm{c}$, where the black line and gray shaded area in Fig.\,\ref{Fig4}c correspond to the blue line and blue shaded area in Fig.\,\ref{Fig4}b, respectively. In other words, a small variation in bacterial dimensions that is natural in an {\it E. coli} suspension is capable for a significant modification of $u_s/R_\mathrm{c}$ that leads to a strongly diversified $TB^*$. Furthermore, for dynamical quantities readily accessible in experiments, our results predict positive correlation between $R_\mathrm{c}$ and $D$ (Fig.\,\ref{Fig4}d). 

\textit{Discussion --} In this study we have quantitatively demonstrated that bacterial dimensions sensitively regulate the translational diffusivity and residence time near a surface, two dynamical properties that are vital to bacterial bio-functions such as infections and nutrients uptake. With the prediction of two linear relations, $\ln{\tau}\propto R_\mathrm{t}/R_\mathrm{b}$ and $D_\mathrm{max}\propto R_\mathrm{t}l/R_\mathrm{b}^2$, a small variation in bacterial dimensions natural to any bacterial suspensions can give rise to strongly diversified surface residence (100-fold change in $\tau$, Fig.\,\ref{Fig3}a) and transport behaviors (10-fold change in $D_\mathrm{max}$, Fig.\,\ref{Fig4}b).  While accurate estimate of $\{l,R_\mathrm{b},R_\mathrm{t}\}$ might be challenging for experiments, we also predict the following correlations between dynamical properties that are readily accessible: anticorrelation between $\ln{\tau}$ and $R_\mathrm{c}$ (Fig.\,\ref{Fig3}b), anticorrelation between $\ln{\tau}$ and $\ln{D}$ (Fig.\,\ref{Fig3}c), and positive correlation between $R_\mathrm{c}$ and $D$ (Fig.\,\ref{Fig4}d).

\parskip = 5pt plus 1pt

\parskip = 5pt plus 1pt

We thank L. S. Luo, M. X. Luo, and M. Stynes for helpful discussions. This work is supported by NSFC No. 12474191, No. 11974038 and No. U2230402. We also acknowledge the computational support from the Beijing Computational Science Research Center.

\bibliography{BacterialEscape}

\end{document}